# Hexagonal boron-carbon fullerene heterostructures; Stable two-dimensional semiconductors with remarkable stiffness, low thermal conductivity and flat bands

Bohayra Mortazavi*,a,b, Yves Rémondc, Hongyuan Fangd, Timon Rabczuke and Xiaoying Zhuang**,a,f

aChair of Computational Science and Simulation Technology, Department of Mathematics and Physics, Leibniz Universität Hannover, Appelstraße 11,30167 Hannover, Germany.
bCluster of Excellence PhoenixD (Photonics, Optics, and Engineering–Innovation Across Disciplines), Gottfried Wilhelm Leibniz Universität Hannover, Hannover, Germany.
cUniversity of Strasbourg, CNRS, ICUBE Laboratory, Strasbourg, France.
dYellow River Laboratory/Underground Engineering Research Institute, Zhengzhou University, Zhengzhou 450001, China.
eInstitute of Structural Mechanics, Bauhaus-Universität Weimar, Marienstr. 15, 99423 Weimar, Germany.
fCollege of Civil Engineering, Department of Geotechnical Engineering, Tongji University, 1239 Siping Road Shanghai, China.

## Abstract

Among exciting recent advances in the field of two-dimensional (2D) materials, the successful fabrications of the $C_{60}$ fullerene networks has been a particularly inspiring accomplishment. Motivated by the recent achievements, herein we explore the stability and physical properties of novel hexagonal boron-carbon fullerene 2D heterostructures, on the basis of already synthesized $B_{40}$ and $C_{36}$ fullerenes. By performing extensive structural minimizations of diverse atomic configurations using the density functional theory method, for the first time, we could successfully detect thermally and dynamically stable boron-carbon fullerene 2D heterostructures. Density functional theory results confirm that the herein predicted 2D networks exhibit very identical semiconducting electronic natures with topological flat bands. Using the machine learning interatomic potentials, we also investigated the mechanical and thermal transport properties. Despite of different bonding architectures, the room temperature lattice thermal conductivity of the predicted nanoporous fullerene heterostructures was found to range between 4 to 10 W/mK. Boron-carbon fullerene heterostructures are predicted to show anisotropic but also remarkable mechanical properties, with tensile strengths and elastic modulus over 8 and 70 GPa, respectively. This study introduces the possibility of developing a novel class of 2D heterostructures based on the fullerene cages, with attractive electronic, thermal and mechanical features.

## 1. Introduction

Fullerenes are originally a group of molecules consisting of carbon atoms, arranged in spherical-like, zero-dimensional (0D) structures, with spatial configuration of hexagonal and pentagonal chains. Fullerene molecules can take various forms, made of different number of atoms arranged in dissimilar chain topologies, making them among the largest molecular systems [1–4]. The first experimental realization of a fullerene molecule was reported in 1994 for the $C_{60}$ lattice [5]. It is nonetheless worth noting that fullerene molecules are not limited to carbon atoms, as demonstrated by the successful production of the full-boron $B_{40}$ counterpart in 2014 [6]. The structural and physical properties of fullerenes make them attractive candidates for a wide range of applications [7–9]. For instance, full-carbon fullerenes can exhibit good electrical conductivity, surpassing that of the copper, which makes them useful in electronics and energy storage systems [8,10]. Additionally, fullerenes have been proven to be promising candidates for pharmaceutical applications [7,9]. The unique chemical, physical, and biological properties of fullerenes make them a versatile and important class of molecules with potential applications in various fields.

One of the latest breakthroughs in the field of fullerenes and 2D nanomaterials is undoubtedly related to the synthesis of the quasi-hexagonal-phase of $C_{60}$ fullerene (qHPC$_{60}$) by Hou *et al.* [11] in 2022. Theoretical studies [4,12–19] have confirmed low thermal conductivity, decent tensile strength and stability and semiconducting character of the qHPC$_{60}$ monolayer. Following the pioneering work by Hou *et al.* [11], two other research groups have also succeeded in creating 2D networks of $C_{60}$ fullerene [20,21]. These very recent experimental successes [11,20,21], confirm synthesizability of diverse fullerene-based 2D networks based on the prominent $C_{60}$ fullerene [4]. In fact, in our recent study [22] by considering the experimentally available $B_{40}$ fullerene [6] and screening of diverse atomic configurations, we could successfully detect the first thermally and dynamically stable full-boron fullerene network with an isotropic structure and metallic electronic character.

Taking into account the exceptional ability of boron and carbon atoms to form strong covalent interactions, an interesting question that arises here is the possibility of realizing stable boron-carbon fullerene coplanar heterostructures, as those experimentally reported for their atomic configurations in $BC_3$ [23] or borophene-graphene heterostructures [24]. In order to address the aforementioned curiosity, herein for the first time we conduct theoretical calculations for exploring the stability and key physical characteristics of the hexagonal boron-carbon fullerene 2D heterostructures, on the basis of $B_{40}$ [6] and $C_{36}$ [25] fullerene molecules. We considered the $B_{40}$ [6] and $C_{36}$ [25] fullerene molecules, not only because they have been experimentally fabricated but also because of their close atomic sizes. Worth noting that $C_{36}$ fullerene can appear with 2d and 6h symmetries [25], however only the 6h configuration was considered in this study since it is experimentally producible [25]. By performing the energy minimization of several hundred randomly generated atomic configurations arranged in hexagonal lattices, for the first time several stable boron-carbon fullerene networks were detected. Density functional theory (DFT) calculations reveal that the boron-carbon fullerene heterostructures can be fabricated and may show semiconducting electronic nature with flat topological bands.

## 2. Computational methods

VASP package [26,27] was used to perform DFT calculations within PBE/GGA functional, D3 [28] vdW dispersion correction and cutoff energy of 500 eV. Optimization of lattices dimensions and ionic positions were accomplished using the conjugate gradient method employing 3×3×1 Monkhorst-Pack [29] k-point grid, with energy and force convergence criteria of $10^{-5}$ eV and 0.005 eV/Å, respectively. To prevent vdW interactions with neighbors, a vacuum spacing of around 15 Å was employed. Ab-initio molecular dynamics (AIMD) simulations were carried out at 500 K with a fixed time step of 1 fs to examine the thermal stability and prepare the training dataset. Moment tensor potentials (MTPs) [30] were fitted to investigate the phononic and mechanical properties using the MLIP package [31]. To prepare the training dataset for MTP, AIMD calculations were performed over unitcells within PBE/GGA, D3 [28] dispersion correction, and 2×2×1 k-point grid for unstrained and biaxially strained systems, using the same method as that of our latest studies [4,32–34]. We generated around 9600 configurations, which are given in the data availability section. A MTP with a cutoff distance of 3.5 Å was fitted using the two-step passive training approach [22] as that of our recent studies [32,34]. In Table S1 of the supporting information document, the errors of the fitted MTP over the training and complete AIMD dataset are compared, which confirms the outstanding accuracy of the developed interatomic model. The phonon dispersion relations were obtained using the developed MTP [35], over 2×2×1 supercells with the aid of PHONOPY package [36]. LAMMPS package [37] was used to study the thermal and mechanical properties using a fixed time step of 0.5 fs within the classical molecular dynamics calculations. Quasi-static uniaxial loading were applied to obtain the mechanical properties [22]. Classical equilibrium molecular dynamics (EMD) simulations were conducted to evaluate the thermal conductivity of the boron-carbon fullerene heterostructures.

## 3. Results and discussions

We will first outline the developed strategy for reconstructing the boron-carbon fullerene heterostructures. The developed approach involved the energy minimization of the isolated $B_{40}$ [6] and $C_{36}$ [25] cages, which were then utilized to generate 2D structures. To ensure complete arbitrariness in the orientation of the molecules, three random rotations were applied with respect to the center of masses of the both molecules. Next, the two molecules were placed in the hexagonal-like unitcells, in which the simulation box dimensions and positions of the molecules were altered until preliminary carbon-boron bonds appear inside and with the periodic images, constructing the initial 2D lattices. Because of the excessive computational costs of DFT calculations, we considered only the hexagonal-like primitive cells, due to the fact that it is the smallest possible configuration to construct periodic coplanar heterostructures made of dissimilar fullerene molecules. An energy minimization with DFT was then subsequently performed on the generated lattices to find the minimum energy configuration. In this work over 400 configurations were fabricated and geometrically optimized, and their total energies were calculated. The lattices with the lowest energies were identified by comparing the energies of different structures. We successfully predicted several $B_{40}$-$C_{36}$ fullerene heterostructures, with the first six lowest structures selected as illustrated in Fig. 1, along with the isolated $B_{40}$ and $C_{36}$ cages, which are also included in the data availability section. As we considered only single $B_{40}$ and $C_{36}$ cages arranged in the hexagonal-like unitcells, we were unable to explore complete potential energy surface [38]. Therefore, in order to detect more extensive stable structures, it is necessary not only to include more randomly molecules in the primitive cell, but also employ the global optimization methods [39–42] to find most stable configurations for the fullerene molecules. Additionally, by continuing the calculations for more random configurations, other low energy systems could be also generated. It is important to note that although the bonding architectures differ among the generated systems, their total energies exhibit marginal differences, only around 0.02 eV/atom. If we take the energy of the isolated $B_{40}$ and $C_{36}$ cages as the reference values, all generated structures exhibit negative energies, which confirm that they are energetically stable. Nonetheless, if we consider the most stable 2D forms of the $B_{40}$ and $C_{36}$ fullerene networks as the reference systems, the respective energy of the most stable system predicted in this work (Fig. 1f) changes from -0.081 to +0.035 eV/atom. This means that there exists higher probability that in the experimentally fabricated samples, the heterostructures appear with domains of full carbon and boron fullerene networks, similar to those observed in the coplanar borophene-graphene [24] or hBN-graphene heterostructures [43]. In order to study the bonding mechanism in the boron-carbon fullerene heterostructures, in Fig. 1 the electron localization function (ELF) [44] with a constant isosurface value of 0.75 are illustrated for every structure. ELF varies from 0 to 1 in every point in the space and close values to unity reveal the perfect localization of electrons. As observable and consistent for all constructed $B_{40}$-$C_{36}$ fullerene heterostructures, large ELF values over 0.75 around the center of C-C, B-C and the majority of B-B bonds, indicate the dominance of the covalent interactions in these novel nanosheets.

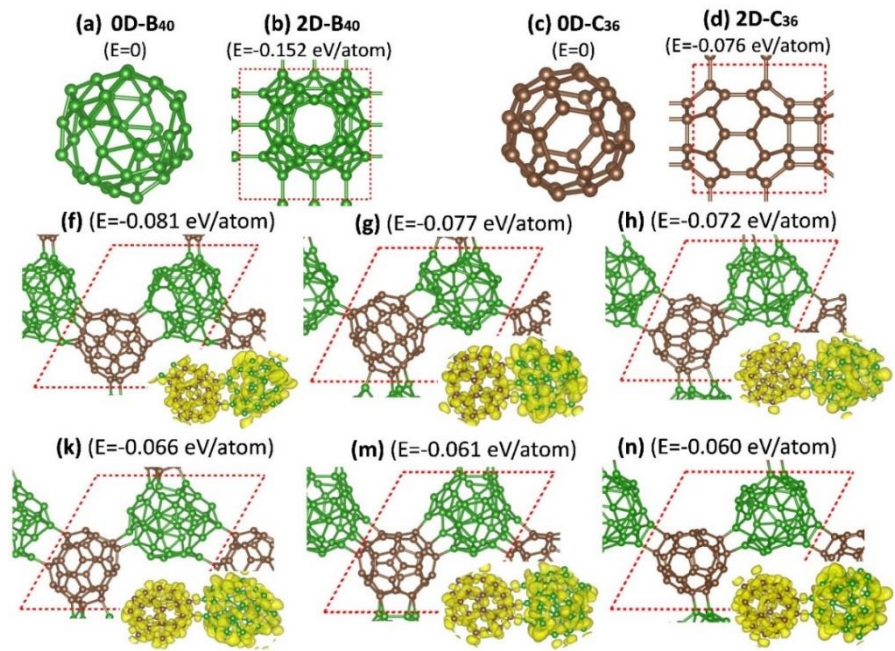

**Fig. 1**. Top views for the molecular geometries of the (a, b) $B_{40}$ and (c, d) $C_{36}$ (with 6h symmetry) 0D and 2D fullerenes. (g-n) Predicted $B_{40}$-$C_{36}$ fullerene heterostructures, ranked according to their total energy with respect to that of the isolated $B_{40}$ and $C_{36}$ molecules. The insets for every lattice show the 3D view for the electronic localization function with an isosurface value of 0.75.

Having analyzed the structural and energetic characteristics of the $B_{40}$-$C_{36}$ fullerene heterostructures, we now shift our attention to the investigation of the dynamical and thermal stability. In Fig. 2, the phonon dispersion curves of the predicted lattices based on the developed MTP is presented. For the all heterostructures, it is observable that none of the three acoustic branches exhibit imaginary frequencies, indicating their desired dynamical stability. Only for the second and third $B_{40}$-$C_{36}$ structures, an optical mode appears with an imaginary frequency. Similar imaginary frequencies have been also observed for the experimentally realized $C_3N_4$ [35] monolayers, which reveal marginal inaccuracies in the geometry optimization. The corresponding group velocities for the phonon branches for every structure is also calculated and presented in insets of Fig. 2. AIMD results at 500 K shown in Fig. S2 confirm that these systems show suitable thermal stability. The AIMD and phonon spectra results demonstrate the acceptable thermal and dynamical stability of the boron-carbon fullerene heterostructures.

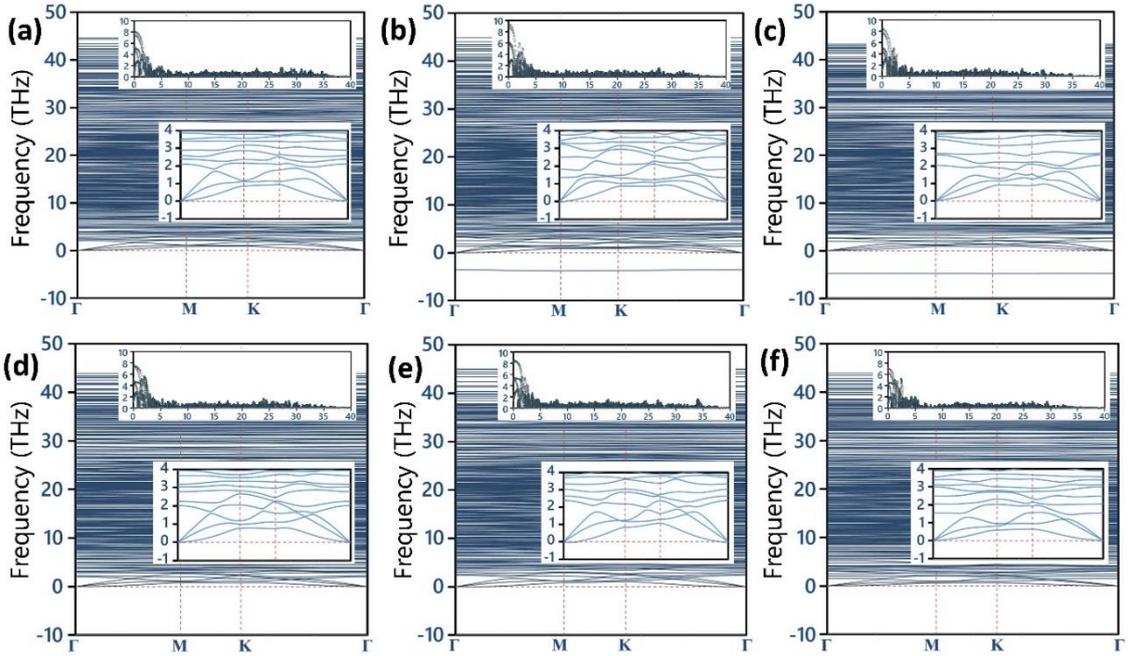

Fig. 2. Phonon dispersion relations of the predicted $B_{40}$-$C_{36}$ fullerene heterostructures along the Γ-M-K-Γ path of the hexagonal structures (find Fig. S1 for examples of dispersions along more complete paths). Upper and lower insets for every panel illustrate the phonons' group velocity in the km/s unit and zoomed view of the low frequency modes in the THz unit, respectively.

We next investigate the lattice thermal conductivity of the $B_{40}$-$C_{36}$ structures at room temperature by conducting the EMD simulations on the basis of the trained MTP using supercell systems consisting of 2736 atoms. In this approach, the structures underwent an initial relaxation process at room temperature using the Berendsen thermostat method for a duration of 100 ps. Subsequently, prior to assessing the thermal conductivity using the EMD method, we conducted constant energy (NVE) simulations for another 50 ps. The EMD method relies on establishing a connection between the ensemble average of the heat current auto-correlation function (HCACF) and the thermal conductivity ($\kappa$) through the Green-Kubo expression:

$$\kappa_a = \frac{1}{VK_B T^2} \int_0^\infty \langle J_a(t) J_a(0) \rangle \, dt \qquad (1)$$

where $a$ denotes the three Cartesian coordinates, $V$ and $T$ are the volume and temperature of the system, respectively and $K_B$ is the Boltzmann's constant. We assumed a fixed thickness of 6.06 Å for these systems, on the basis of that for the $C_{36}$ fullerene 2D network [4]. The auto-correlation functions of the heat current $\langle J_a(t) J_a(0) \rangle$ can be calculated using the heat current $\vec{J(t)}$ as expressed by [37]:

$$\vec{J(t)} = \sum_i \left( e_i \vec{v_i} + \frac{1}{2} \sum_{i<j} (\vec{f_{ij}} \cdot (\vec{v_i} + \vec{v_j})) \vec{r_{ij}} \right) \qquad (2)$$

here, $e_i$ and $v_i$ are respectively the total energy and velocities of atom i. In addition, $f_{ij}$ and $r_{ij}$ are the interatomic force and position vector between atoms i and j, respectively. During the NVE calculations, we recorded the heat current values along the planar direction to compute the heat current auto-correlation functions (HCACFs). Ten separate simulations were conducted with uncorrelated initial velocities, and the resulting HCACFs were averaged to determine the effective thermal conductivity using Eq. 1. Because of the very similar phononic characteristics of the constructed $B_{40}$-$C_{36}$ fullerene heterostructures, in this case we only consider the third and sixth lattices as the representative structures for the analysis of thermal transport. Worth reminding that the sixth structure exhibits the lowest maximum group velocity, which can be useful to find the lower bound of thermal conductivity. In Fig. 3, the calculated room temperature lattice thermal conductivity of the two considered lattices along the two in-plane directions as a function of correlation time are illustrated. As it can be seen, the averaged thermal conductivity tends to converge at correlation times longer than 20 ps. The sixth $B_{40}$-$C_{36}$ fullerene heterostructure is found to exhibit a convincingly isotropic averaged thermal conductivity around 4.4±0.3 W/mK at room temperature. On the other hand, the third fullerene heterostructure presents a slightly anisotropic averaged thermal conductivity at room temperature, 7.3±0.5 and 8.6±0.5 W/mK, along the two planar directions. As expected, the higher phonon group velocity of the third structure resulted in enhanced thermal transport. Consistent with the conclusion of our recent study on the $C_{36}$ fullerene network [4], the room temperature lattice thermal conductivity of the carbon-boron fullerene heterostructures are also expected to be in the order of 10 W/mK.

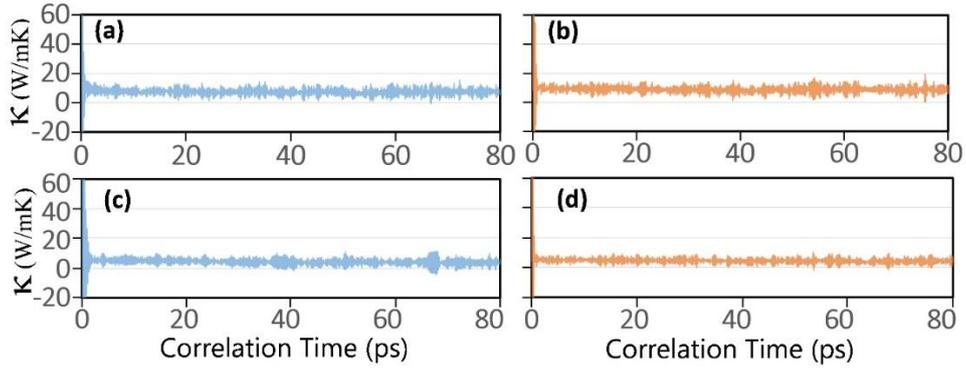

Fig. 3. EMD-based results for the calculated room temperature lattice thermal conductivity of the (a, b) third (c, d) sixth single-layer $B_{40}$-$C_{36}$ fullerene heterostructures as a function of correlation time along the (a, c) horizontal and (b, d) vertical directions, shown in Fig. 1.

After the analysis of phononic properties, the developed molecular dynamics models are employed to investigate the anisotropic mechanical responses of the $B_{40}$-$C_{36}$ fullerene heterostructures. Fig. 4 shows the predicted direction-dependent uniaxial stress-strain responses of the six $B_{40}$-$C_{36}$ fullerene heterostructures simulated at 1 K using 4×4×1 supercells. In the presented results, we considered the real volumes of the deformed lattices in evaluating the stress value, assuming a fixed thickness of 6.06 Å as that the $C_{36}$ network [4] during the uniaxial loading. The $C_{11}$, $C_{12}$ and $C_{22}$ elastic constants for $B_{40}$-$C_{36}$ lattices are also summarized in Fig. 4 results. The present results reveal highly anisotropic elasticity and tensile behavior in the predicted 2D networks. It is moreover noticeable that the predicted fullerene-based networks generally show irregular stress-strain curves, particularly after exceeding the ultimate tensile strength point, showing series of stress rises and consequent drops. Close to the ground state, the tensile strength of the $B_{40}$-$C_{36}$ fullerene heterostructures are estimated to be within 8-15 GPa, which are remarkably high for nanoporous and light-weight structures. In accordance with the earlier analysis of the phononic properties, the sixth structure not only shows the lowest elastic and tensile strength values, but also exhibits the most isotropic behavior. In the aforementioned system, individual $B_{40}$ and $C_{36}$ fullerene cages are connected with only two B-C bonds along the two planar directions.

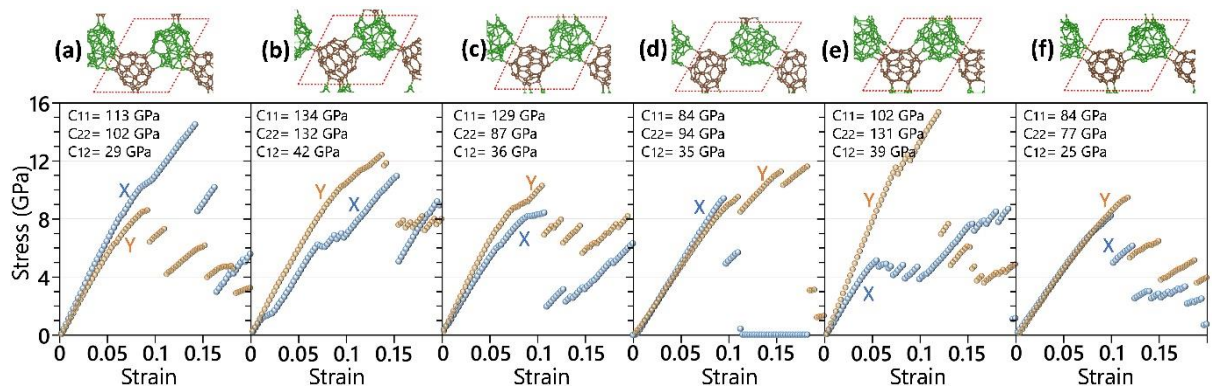

**Fig. 4**, Uniaxial stress-strain responses of the $B_{40}$-$C_{36}$ fullerene heterostructures by MTP-based molecular dynamics simulations conducted at 1 K. The directions 1 and 2 for elastic constants refer to the horizontal (X) and vertical (Y) directions, respectively.

As discussed previously, the formation of covalent bonding between carbon and boron atoms in the adjacent fullerene cages, reinforce the mechanical strength of these systems. In comparison with the full-carbon fullerene networks, which exhibit brittle failure behaviors [4,12–17], the presented stress-strain curves shown in Fig. 4 reveal ductile-like behavior along the carbon-boron counterparts. In order to analyze the underlying mechanism responsible for ductile behavior, in Fig. 5 the deformed structures are illustrated. As it is clear, while carbon cages generally tend to keep their spherical geometry under the mechanical loading, the boron counterparts undergo remarkable deformations and deflections. In addition, the formation of boron atoms chains results in interconnecting of fullerene cages at higher strain levels, prohibiting the sudden brittle failure. In these systems, the deformation initially proceeds with the bond elongation, but continues with combination of structural deflection and bond breakages, resulting in complex tensile behavior. As it is clear, boron atoms because of their unique ability for taking various forms and undergoing large deflection without bond breakages, are the main contributors to the irregular stress-strain curves and ductile-like behaviors in the fullerene heterostructures.

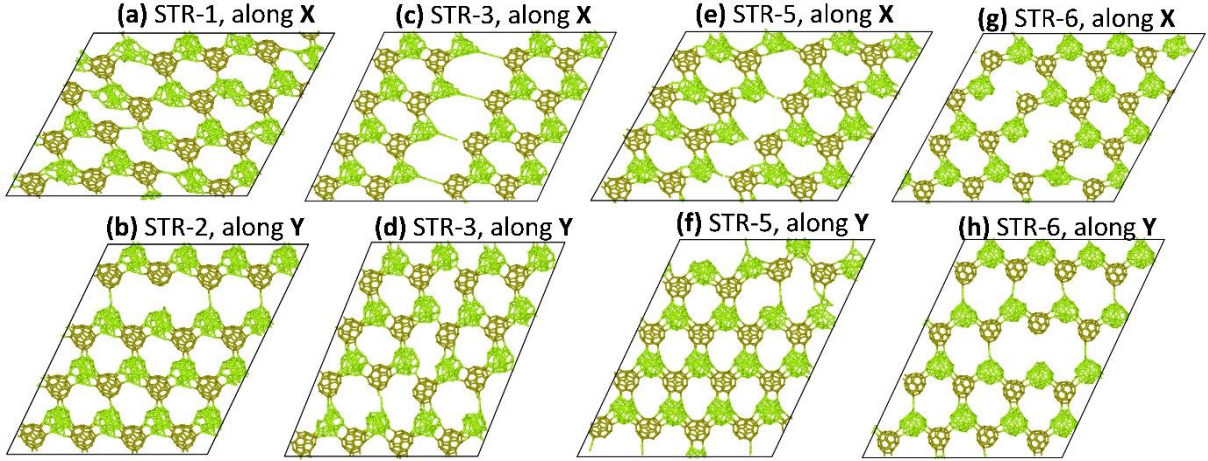

Fig. 5, Deformed $B_{40}$-$C_{36}$ fullerene heterostructures under uniaxial loading along the (first row) X and (second row) Y directions for various structures (STR).

Last but certainly not least, we briefly investigate electronic features of the $B_{40}$-$C_{36}$ fullerene heterostructures on the basis of the PBE/GGA functional using the VASPKIT [45] to find the symmetry points, which are illustrated in Fig. 6. As it is clear, different $B_{40}$-$C_{36}$ fullerene heterostructures despite their completely dissimilar bonding architectures, exhibit semiconducting electronic nature with rather flat dispersions for the both valance and conduction bands, implying low carrier mobilities in these systems. Interestingly, the fourth structure is a direct gap semiconductor in the vicinity of the X point, which shows the narrowest band gap of 0.13 eV and also exhibits the sharpest conduction band minimum dispersion. The second and third $B_{40}$-$C_{36}$ fullerene networks, both yield a band gap of 0.61 eV, with considerably flat dispersions for the valance and conduction bands. In the data availability section, the band gap value, positions of the valance band maximum and conduction band minimum and complete PBE/GGA electronic band structure for every structure are given. The presented results indicate that the $B_{40}$-$C_{36}$ fullerene heterostructures are generally narrow-gap semiconductors with flat valance and conduction bands near the Fermi energy. The observed flat bands indicate large density of states, which amplifies the effects of interactions in these systems. In the $B_{40}$-$C_{36}$ fullerene networks, even weak interactions can yield significant contribution to the electronic and optical properties, lifting the near degeneracy between occupying states near the Fermi energy, which are highly appealing for further theoretical explorations, particularly concerning the superconducting transition. We should nonetheless remind that the PBE/GGA method scientifically underestimates the band gap of the semiconducting materials, and therefore with more precise HSE06 [46] or advanced GW [47] methods the estimated band gaps will increase. Moreover, while the mechanical and thermal conductivity generally do not show considerable dependency on the substrate, the electronic properties can be substantially affected by the orientation and chemical composition of the substrate.

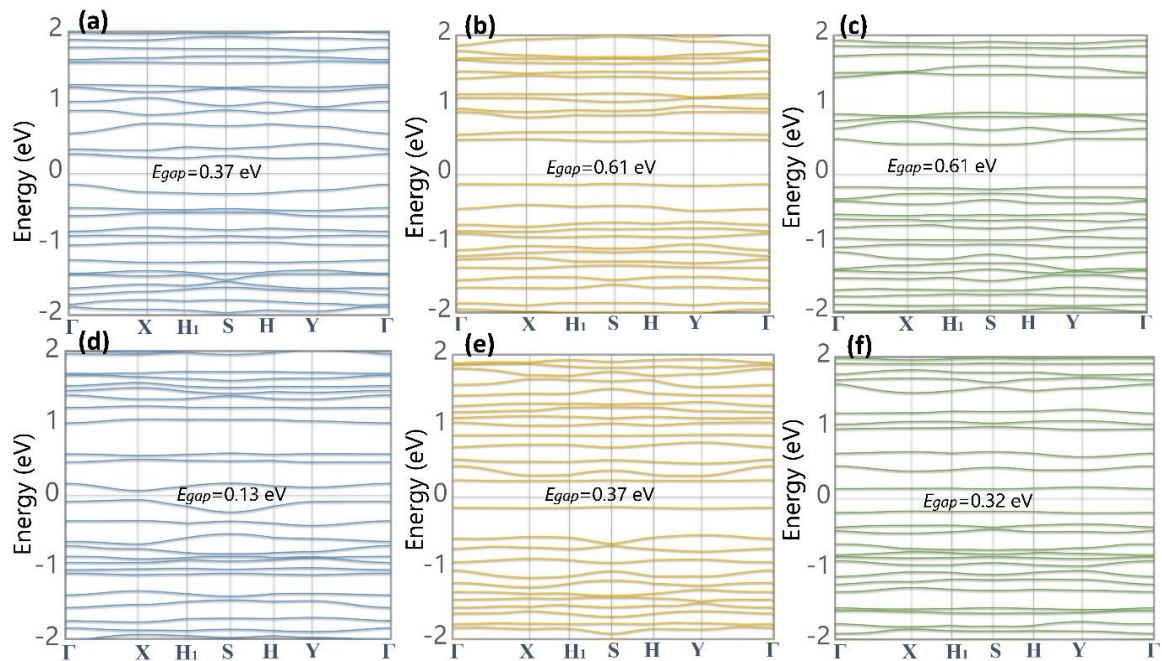

Fig. 6. Electronic band structures of the single-layer $B_{40}$-$C_{36}$ fullerene heterostructures with the PBE/GGA functional. The Fermi energy is set to 0 eV.

## 4. Concluding remarks

In the present investigation, for the first time we explored the stability and physical characteristics of the coplanar 2D heterostructures composed of boron and carbon fullerenes. As a practical case study, we consider the previously synthesized $B_{40}$ and $C_{36}$ fullerenes as the consisting materials. By conducting structural minimization over several hundred randomly generated hexagonal 2D structures, we could for the first time predict stable 2D heterostructures made of boron and carbon fullerenes. Nonetheless, it was concluded that in experimentally fabricated samples, the heterostructures most probably will appear with domains of full carbon and boron fullerene networks, tightly connected by strong boron-carbon bonds. According to molecular dynamics simulations results on the basis of machine learning interatomic potentials, the room temperature lattice thermal conductivity of the boron-carbon fullerene heterostructures are expected to be within 4 to 10 W/mK. Moreover, these novel 2D networks are predicted to show anisotropic but also remarkable mechanical properties, with tensile strengths and elastic modulus over 8 and 70 GPa, respectively. Due to the presence of the boron clusters, the deformation of $B_{40}$-$C_{36}$ fullerene heterostructures is observed to proceeds with combinations of bond elongation, structural deflection and bond breakages, resulting in irregular stress-strain curves and ductile-like failure mechanism. Interestingly, it is shown that different $B_{40}$-$C_{36}$ fullerene heterostructures with completely dissimilar bonding architectures, generally exhibit semiconducting electronic nature with rather flat dispersions for the both valance and conduction bands, with PBE/GGA gaps ranging between 0.13 to 0.61 eV. Our theoretical investigation, highlights the feasibility of the synthesis of the boron and carbon fullerene heterostructures, which are expected to show decent stability and mechanical stiffness, low thermal conductivity and semiconducting electronic nature with flat bands, highly attractive to explore new physics and chemistry. Resources

### Declaration of competing interest
The authors declare that they have no known competing financial interests or personal relationships that could have appeared to influence the work reported in this paper.

### Data availability
Please find: https://data.mendeley.com/datasets/tgszs66spj/1 for the related data for this work, which includes: energy minimized atomic lattices in the VASP POSCAR format, summary of electronic properties, complete AIMD dataset for the training and the fitted MTP (atomic types 0 and 1 represent C and B, respectively) for the evaluation of mechanical response and thermal conductivity.

### Acknowledgment
B. M. and X. Z. appreciate the funding by the Deutsche Forschungsgemeinschaft (DFG, German Research Foundation) under Germany's Excellence Strategy within the Cluster of Excellence PhoenixD (EXC 2122, Project ID 390833453). B. M and T. R. are greatly thankful to the VEGAS cluster at Bauhaus University of Weimar for providing the computational resources.